# Big Data: Opportunities and Privacy Challenges

## Hervais Simo

Fraunhofer-Institut für Sichere Informationstechnologie, Darmstadt, Germany

## Table of Content

**Abstract.**
Recent advances in data collection and computational statistics coupled with increases in computer processing power, along with the plunging costs of storage are making technologies to effectively analyze large sets of heterogeneous data ubiquitous. Applying such technologies (often referred to as big data technologies) to an ever growing number and variety of internal and external data sources, businesses and institutions can discover hidden correlations between data items, and extract actionable insights needed for innovation and economic growth. While on one hand big data technologies yield great promises, on the other hand, they raise critical security, privacy, and ethical issues, which if left unaddressed may become significant barriers to the fulfillment of expected opportunities and long-term success of big data. In this paper, we discuss the benefits of big data to individuals and society at large, focusing on seven key use cases: Big data for business optimization and customer analytics, big data and science, big data and health care, big data and finance, big data and the emerging energy distribution systems, big/open data as enablers of openness and efficiency in government, and big data security. In addition to benefits and opportunities, we discuss the security, privacy, and ethical issues at stake.

**Keywords.** Big Data, Opportunities, Privacy, Informational Self-determination

## Introduction

The volume and variety of data produced by and about individuals, things or the interactions between them have exploded over the last few years. Such data can be replicated at low cost and is typically stored in searchable databases which are publicly (or at least easily) accessible over the Internet. According to recent IBM estimates, 2.5 billion Gigabytes of data are created everyday around the globe, and the creation rate is growing continuously.[1] McKinsey estimates that the amount of digital content on the Internet is expected to grow by 44 times to 2020, at an annual growth rate of 40%.[2] This trend describes a phenomenon broadly known as the emergence of big data. The big data phenomenon itself is in part being enabled by the rising popularity of Web 2.0 (esp. online social networks) applications, the low cost of computation and storage, the rapid emergence of new computing paradigms such as cloud computing, breakthrough innovations in the field of data mining and artificial intelligence, combined with the wide availability of sensor-equipped and Internet-compatible mobile devices.

Although there is still no commonly agreed upon definition of "big data", the term is often used to describe the exponential growth and availability, as well as the variety of data (of different format, nature, or origin) and speed at which it is produced and transferred. Key bodies such as the U.S. National Institute of Standards and Technology (NIST) and the research analyst Gartner have promoted a definition encompassing three dimensions: Volume (i.e. the amount of data), Velocity (i.e. the speed of data), and Variety (i.e. the array of data types and sources).[3] Recently, technology giants, including IBM [4], have extended the relatively standard 3Vs definition of big data to include another V: Veracity (i.e., the quality and accuracy of data). Note that the flood of data and content contributing to the big data phenomenon does not only include data originally created, stored and processed for a certain purpose, but also information which is a by-product of other electronic transactions. Furthermore, note that big data is different from traditional data warehousing and types of business intelligence analysis that have been around for a while. Unlike in traditional data management scenarios, a large part of big data is unstructured and raw (a.k.a. "grey data") data that is generated with greater velocity than ever before. Examples of such unstructured and raw data include email messages, images, audio and video files, and GPS coordinates. Relying on high-performance, low-cost storage infrastructures and powerful data mining techniques and statistical correlation algorithms, data analysts are increasingly able to extract complex patterns, discover correlations and cull valuable information from compilations of real-time and cross-domain data. Examples of big data sources include organizations' Intranets and online government directories; the ever-growing mountains of search logs, clickstream, and (mobile) network traces; records of users' online social interactions; ubiquitous cyber-physical systems such as smart energy distribution systems, intelligent transport systems (ITS) that relies on cars which are becoming smarter than ever, and intelligent home systems that among other things integrate different home entertainment platforms, interconnect a variety of home appliances ranging from thermostats to home-security devices, and support face and emotion recognition applications through various motion-detection technologies.

---

[1] http://www-01.ibm.com/software/data/bigdata/
[2] Manyika, James, et al. "Big data: The next frontier for innovation, competition, and productivity." (2011).
[3] Ward, Jonathan Stuart, and Adam Barker. "Undefined By Data: A Survey of Big Data Definitions." arXiv preprint arXiv:1309.5821 (2013).
[4] IBM, The Four V's of Big Data. http://www.ibmbigdatahub.com/infographic/four-vs-big-data

This trend is supporting the rise of a broad variety of services that are highly customized to various aspect of our life, and hold great social and economic potential.[5][6] Big data analysts are indeed able to apply smart algorithms and artificial intelligence to large sets of data can discover hidden insights relevant in various scenarios, from data-driven decision optimization (e.g., optimization of police proactive tactical decision making to reduce crime), and healthcare (e.g., patients' risk for a certain rare diseases and tracking the spread of influenza viruses), to improving our understanding of human behaviour in certain socio-technical environments. However, as data is increasingly viewed as a commodity and new form of currency, the emergence of such huge amounts of aggregated data and their linkability to other datasets clearly introduce a whole new set of privacy challenges. The increasing ubiquitousness of large-scale data storage, big data analytics, and automated decision-making impacting critical aspects of peoples' lives based on opaque algorithms raises concerns over threats to peoples's right to informational self-determination, unfair discrimination, and other prejudicial outcomes.

## 1. The Power and the Promises of Big Data

As institutions and businesses are becoming inherently data driven, a widespread deployment of effective big data technologies can significantly contribute to innovation and enable increased productivity and economic growth, from which not only businesses but society at large would benefit. [7][8][9][10] This section discusses examples of technologies and application scenarios, illustrating the promise and potential big data. Some of the scenarios are already reality while others are expected to be implemented in the (near) future. At this point, we do not claim that our set of examples is exhaustive, as continued big data market growth is expected and new big data scenarios are likely to evolve with the technology. We refer the interested readers to [11] [12] [13] [14] for additional scenarios of big data technologies.

---

[5] Bollier, David, and Charles M. Firestone. The promise and peril of big data.Washington, DC, USA: Aspen Institute, Communications and Society Program, 2010.

[6] Mayer-Schonberger, V., and K. Cukier. "Big data: A revolution that will change how we live, work and think." (2013).

[7] Manyika, James, et al. "Big data: The next frontier for innovation, competition, and productivity." (2011).

[8] Bollier, David, and Charles M. Firestone. The promise and peril of big data.Washington, DC, USA: Aspen Institute, Communications and Society Program, 2010.

[9] HO, Diem, et al. Unleashing the Potential of Big Data. Organizational Design Community, 2013

[10] Mayer-Schonberger, V., and K. Cukier. "Big data: A revolution that will change how we live, work and think." (2013).

## 1.1. Big Data for Business Optimization and Customer Analytics

For business leaders and marketers, big data could mean the key to a new era of business intelligence and personalized business service delivery. [15][16][17] By relying on the integration of advanced analytics and modern data warehouse platforms, business analysts can extract and visualize hidden patterns and meaningful insights from various internal and external data sources. Leveraging this knowledge, they can add intelligence to their processes, improve operational efficiency, and as a result gain competitive advantages. [18][19] Indeed, by cross-linking complex, heterogeneous, and large data sets, i.e., data from companies internal sources and the growing torrent of heterogeneous data available externally, business analysts may be able, among other things, to optimize their marketing and advertising strategies, gain real-time insight into their customers' needs, usage, and buying patterns, and possibly identify emerging (product/market) trends early on. Many companies, especially (online) retailers, are already applying big data techniques to their vast databases of consumer purchase histories, transactional information and inventory data to i) gain a better understanding of their customers, ii) provide potential and current customers with personalized products, services, and recommendations, and iii) predict shifts in demand. Relying on similar analytic techniques, sales and marketing professionals may be able in the not too distant future to leverage the mountain of data that customers creates when using mobile/smart devices to access online services, when making online purchases with electronic cards, or when sharing their whereabouts and intimate thoughts on OSNs to target the right consumer at the right time with the right message. [20] According to a 2011 survey carried out by McKinsey [21], early adopters of the Big data technologies could increase their operating margins by 60%. A recent survey by the German market research institute *Gesellschaft für Konsumforschung* (Gfk), found out that 86% of marketers consider big data as a "game-changer", with 62% saying that their role has already changed as a result of it. [22]

---

[11] Mayer-Schnberger, V. and Cukier, K. (2013) Big Data: A Revolution That Will Transform How We Live, Work and Think. John Murray.
[12] Bollier, David, and CharlesM. Firestone. The promise and peril of big data.Washington, DC, USA: Aspen
Institute, Communications and Society Program, 2010.
[13] Manyika, James, et al. "Big data: The next frontier for innovation, competition, and productivity." (2011).
[14] The US President's Council of Advisors on Science and Technology
("PCAST"). "Big Data and Privacy: A Technological Perspective." May 1, 2014,
http://www.whitehouse.gov/sites/default/files/microsites/ostp/PCAST/pcast_big_data_and_privacy_-_may_2014.pdf
[15] Bitkom: Big Data im Praxiseinsatz: Szenarien, Beispiele, Effekte. Bitkom Arbeitskreis Big Data, 2012.
[16] http://www.gfk.com/uk/documents/press-releases/gfk%20release%20for%20changing%20ad%20summit.pdf
[17] Manyika, James, et al. "Big data: The next frontier for innovation, competition, and productivity." (2011).
[18] McAfee, Andrew, and Erik Brynjolfsson. "Big data: the management revolution." Harvard business review
90.10 (2012): 60-68.
[19] Organizational Data Mining: Leveraging Enterprise Data Resources for Optimal Performance
http://books.google.de/books?id=EXh4GqN27L0C&hl=de&source=gbs_navlinks_s
[20] Brown, Brad, Michael Chui, and James Manyika. "Are you ready for the era of big data." McKinsey
Quarterly 4 (2011): 24-35.
[21] Manyika, James, et al. "Big data: The next frontier for innovation, competition, and productivity." (2011).
[22] http://www.gfk.com/uk/documents/press-releases/gfk%20release%20for%20changing%20ad%20summit.pdf

## 1.2. Big Data and Science

Big data has the potential to change science as we know it. Progresses in the last decade in the fields of high-performance computer simulation and complex real-time analytics paired with the rapidly increasing volume and heterogeneity of data from various sources (incl. Web browsing/searching records, genomic, health and medical records, earth observation systems, surveillance video, and sensor, wireless and mobile networks) are shaping the vision of a data-intensive science. [23, 24] The vision of a data-intensive science describes a new approach to the pursuit of scientific exploration and discovery, which leverages an ever-growing amount of research data and thus requires new computing, simulation, and data management tools and techniques. The approach promises to integrate life, physical, and social sciences and covers application domains ranging from computational earth and environmental sciences, genomics, to computational social science. The hope being that data-intensive science would enable mankind to better understand and address some of its most pressing challenges: global warming; efficient supply and use of cleaner energy resources; pandemics and global health monitoring among others. To take the example of computational social science [25]: In the recent years, social scientists have begun collecting and analyzing large volumes of data from sources that were barely imaginable a decade ago. Online social network platforms like Facebook and Twitter, and emerging applications such as participatory sensing, two of such sources, allow for the mass-scale collection and sharing of details about peoples' behaviour, as well as the nature and strength of the interactions between individuals and communities in on- and offline environments. It has been demonstrated that the complex and heterogeneous data from these environments can be leveraged alongside statistical techniques to gain insights into online sociological phenomenenon.[26][27] In particular, the social character of these environments and the nature of the data being collected and analysed enable an interpretation of the sociological phenomenon on both a micro level focusing, for example on an individual's influence, and on a macro-level, for example uncovering behavioural patterns of groups of people.

## 1.3 Big Data is Reshaping Medicine and Health Care

As almost all aspects of healthcare (including public health monitoring, healthcare delivery and research) become more and more dependent on information technology, stakeholders in the healthcare industry and health economics are increasingly able to collect, process and share various types of data systematically, including person-related biological samples, medical imaging data, patient claims, prescriptions, clinical notes, and other medical statistics. As a matter of fact, collecting, processing, and sharing this kind of data is almost as old as clinical medicine. What makes this a greater topic of interest in the big data age, however, is that healthcare analysts and practitioners can now i) combine traditional health data with external data - demographic, behavioural and medical/fitness related sensor data - in order to glean insights into human activities and relationships, and then ii) leverage these insights to improve medical research, discover and monitor otherwise invisible health pat-terns in a large portion of the population, or to provide new innovative personalized medical products and services. (cf. [28][29]) For instance, big data holds the promise of advanced statistical methods that can help geneticists and drug manufacturers correlate large sets of genomic and clinical trial data with streaming data from the Web and government censuses in order to understand better how inherited genetic variants contribute to certain genetic diseases or predispositions to (rare) diseases, and accurately perform drug tests. The collection and subsequent aggregation of behavioural/demographic data about current and would-be patients using traditional clinical and health data through data infrastructures such as ELIXIR[30] may empower clinicians as they seek to improve their ability to diagnose, tailor medical treatment to patients unique genetic profile, and improve the overall standards of care accordingly. Relying on similar big-data analytics software and tools, healthcare authorities may be able in the future to combine epidemiologic methods with morbidity and mortality statistics that they have accumulated over the years in order to gain a better understanding of (rare) disease propagation patterns and/or reassess their disaster recovery plan activities. Furthermore, by applying predictive analytics and simulation to healthcare data, healthcare authorities may gain insights into or predict the demographic distribution of certain diseases with regards to ethnicity, gender, and geography, and be able to accurately quantify the interplay between the quality of healthcare services accessible in different geographic areas and the government's investment in health care. Companies of various sizes have recently started tapping into the unleashed business potential of electronic data increasingly available in the healthcare industry to offer a variety of personalized medicine and genomics services. A 2013 report [31] argues that more than 200 businesses in the US are developing a variety of innovative software tools and platforms to make better use of available healthcare information. McKinsey [32][33] predicts that biotech startups and other players in the health economics who rely on big data analytics would generate a market worth $10 billion by 2020.

---

[23] Tansley, Stewart, and Kristin Michele Tolle, eds. "The fourth paradigm: data-intensive scientific discovery." (2009).
[24] http://spectrum.ieee.org/at-work/innovation/the-coming-data-deluge
[25] Lazer, David, et al. "Life in the network: the coming age of computational social science." Science (New York, NY) 323.5915 (2009): 721.
[26] Predicting the Future With Social Media: http://arxiv.org/pdf/1003.5699v1.pdf
[27] McKelvey, Karissa, et al. "Visualizing communication on social media:Making big data accessible." arXiv preprint arXiv:1202.1367 (2012).

### 1.4. Big Data and Financial Services

Financial institutions are increasingly capitalizing on recent developments in the field of IT and big data related tools. They use these technologies to compile and analyze huge amounts of personal, economic, and financial data, some of which are real-time streams (e.g. those from stock and financial markets), in order to better understand and control the complex compliance challenges and financial risks associated with possible new investments. In recent years, credit agencies and insurers have become eager to capitalize on recent developments in the field of IT and the easy access to social networking data to analyze years of transactional data retrospectively as they seek to detect highly complex patterns, which they can use for fraud detection.[34] Brokerage firms' growing ability to assess large numbers of possible market scenarios, analyze new types and sources of data (e.g. breaking news and weather information, real-time sub-prime market data, social media) may allow them to tease out potentially valuable patterns that would otherwise remain hid-den. They can then use those insights to predict stock market performances and improve trading decisions. One illustrative example of this computational approach to stock market is high-frequency stock trading [35,36]: an emerging form of trading that relies fully on high-speed computers and clever algorithms to make accurate trading decisions at rates measured in the order of milliseconds. Moreover, entire business segments are increasingly relying on big data and complex machine-learning algorithms as they aim to avoid bad lending decisions or managing risks associated with customer payments online. Non-bank lenders, in particular, are expected to apply advanced analytics increasingly on real-time compilations of cross-domain data in order to gain insight into consumer behavior, identify potentially suspicious users activities, and consequently make accurate lending decisions with a precision largely thought impossible just a few years ago.. Summing up, big data driven financial services have the potential to contribute to greater financial inclusion which in turn is vital for archiving inclusive economic growth.

---

[28] Sun, Jimeng, and Chandan K. Reddy. "Big data analytics for healthcare." Proceedings of the 19th ACM SIGKDD international conference on Knowledge discovery and data mining. ACM, 2013.
[29] Groves, Peter, et al. "The big datarevolution in healthcare." McKinsey Quarterly (2013).
[30] http://www.elixir-europe.org/about/
[31] Groves, Peter, et al. "The big datarevolution in healthcare." McKinsey Quarterly (2013).
[32] Groves, Peter, et al. "The big datarevolution in healthcare." McKinsey Quarterly (2013).
[33] Manyika, James, et al. "Big data: The next frontier for innovation, competition, and productivity." (2011).
[34] Surfing for Details: German Agency to Mine Facebook to Assess Creditworthiness
http://www.spiegel.de/international/germany/german-credit-agency-plans-to-analyze-individual-facebook-pages-a-837539.html

## 1.5. Big Data in Emerging Energy Distribution Systems

The energy sector (along with the emerging smart grid applications [37]) is another field witnessing a growing use of data-driven processes and data analytic tools. The increasing deployment of smart meters, intelligent field devices, and other intelligent IT components within modern energy infrastructures is generating a flood of new types of data. [38] A near real-time collection and analysis allow utility companies to make sense of this data to improve the efficiency of power generation, transmission, and distribution, e.g. by being able to predict peak demand at multiple scales, and to model and run higher fidelity simulation of power grids. Furthermore, power utility companies could analyze data about energy use reported by smart meters to detect illicit activities such as electricity theft and fraud, with the potential to bring about major energy and financial savings and environmental benefits. [39,40] Recently, several start-ups have begun to develop new applications that based on behavioral analytics may enable end-users to understand, monitor and actively control their energy usage. According to a recent study by Pike Research, the market for smart grid data analytics is expected to reach a total value of approximately $34 billion from 2012 through 2020.[41]

## 1.6. Big/Open Data - Potential Enablers of Openness and Efficiency in Government

Big data is changing the public sector as well. Over the past 10 years, several governments have kick-started initiatives to publicize large sets of public data, incl. census data, crime statistics, traffic statistics, meteorological data, and healthcare data. The move aims at promoting transparency and government accountability, and achieving efficiency and effectiveness in Government.[42][43][44] Another hope is that an easy access to and use of high value, yet still under-leveraged, government data by private and commercial entities will drive innovations and create a new wave of economic growth. Many businesses and scientists view such freely accessible and searchable mountains of data as gold mines.[45] According to the British government "[...] organizations, and even individuals, can exploit this data in ways which government could not be expected to foresee".[46] The use of advanced data processing and analytical techniques to tap into the ever-growing volume of under-leveraged government data and make relevant information available across different agencies is expected to help governments improve operational efficiency and reduce cost. According to [47], the European public sector is missing out on combined cost savings of around 100 billion per annum by failing to maximize the potential of big data for operational efficiency. Similar benefits and opportunities are expected in other areas/sectors of the society. In the private sector, for instance, one can easily imagine a near future scenario in which start-ups would be able, on a massive scale, to access and correlate publicly available property data (incl. estimated value and location) with crime statistics, aiming at providing theirs customers with personalized recommendations about where to buy or not to buy a property. In the arena of politics, political parties started a few years ago to rely on new data collection and analysis techniques to optimize critical aspects of their campaign operations, including fundraising and the mobilization of grass-root supporters, among others. [48] To that end, they have been applying microtargeting techniques and other analytic methods (cf. [49]) on compilations of public sector data (incl. voter registration records, campaign contributions records, and census data) and data collected through social media, and mobile devices/networks. This trend is expected to continue with the growing need in other (non-)western democracies to better understand and accurately predict voters' attitudes and preferences.

---

[35] Chlistalla, Michael, et al. "High-frequency trading." Deutsche Bank Research (2011): 1-19.
[36] Kirilenko, Andrei, et al. "The flash crash: The impact of high frequency trading on an electronic market."
Manuscript, U of Maryland (2011).
[37] Smart Metering Systems Intelligente Messsysteme: https://www.bsi.bund.de/DE/Themen/SmartMeter/smartmeter_node.html
[38] Framework, N. I. S. T. "Roadmap for Smart Grid Interoperability Standards Release 2.0 [NIST SP-1108R2]." (2012).
[39] Fighting Electricity Theft with Advanced Metering Infrastructure:
http://www.ecitele.com/OurOffering/Industries/IndustriesAssets/energy-fighting-electricity-theft-with-advanced-metering-infrastructure.pdf
[40] Energy Theft in the Advanced Metering Infrastructure. Stephen McLaughlin
[41] Pike Research: Smart Grid Data Analytics, http://www.navigantresearch.com/research/smart-grid-data-analytics
[42] Berlin Internet Institute

## 1.7. Detecting and Fighting (Cyber-) Crime with Big Data

Fighting (cyber-) crime does not only require a retrospective analysis of possible evidences but also accurate predictions about criminals' behaviours and their adaptive reactions to countermeasures. As chief (information) security officers are struggling to monitor and protect their corporate's networks and enterprise systems against increasingly sophisticated and complex security threats, private companies are slowly but surely moving towards adopting big data security analytics tools, i.e., tools that bring advanced data analytics to enterprise IT security. Unlike current security information and event management (SIEM)[50] solutions, big data security analytics tools such as IBM Security Intelligence[51] and Palantir[52] provide the means required to effectively analyze terabytes of (real-time) network events, packet captures, applications' performances and unstructured data from across/outside the organization. By providing means to discover changing patterns of malicious activities hidden deep in large volumes of organizations data, big data security tools can indeed empow-er businesses to better understand if and how they have been attacked.[53] In addition to providing continuous and detailed insight into security risks, big data security analytics tools can help organizations in their regulatory compliance efforts. Moreover, the collected statistics and the possibly inferred knowledge about attacks/ breaches/ criminal activities could then be shared with other organizations across national jurisdictions, for instance in an effort to achieve collaborative network monitoring / collaborative intrusion detection.[54]

---

[43] http://www.bigopendata.eu/full-report/
[44] http://ec.europa.eu/digital-agenda/public-sector-information-raw-data-new-services-and-products
[45] Data, Big. "Big Impact: New Possibilities for International Development." 2013-04-07]. http://www3.weforum, org/docs/WEF-TC-MFS-BigData Big lmpact *Brie f ing*.2012.*pdf* .
[46] The Government of the UK, Further Detail on Open Data Measures in the Autumn Statement 2011 http://www.cabinetoffice.gov.uk/sites/default/files/resources/Further*detailonO penDatameasuresintheAutumnStatement*2011.*pdf*
[47] Manyika/Chui/Brown/Bughin/Dobbs/Roxburgh/Byers 2011.
[48] Nickerson, David W., and Todd Rogers. "Political Campaigns and Big Data." (2013).
[49] Nagourney, Adam. "The'08 campaign: sea change for politics as we know it." The New York Times (2008):1.
[50] http://www.gartner.com/it-glossary/security-information-and-event-management-siem
[51] http://www-03.ibm.com/security/solution/intelligence-big-data/
[52] http://www.palantir.com/solutions/cyber/

Another transformative potential of big data security analytics lies in its power to satisfy the ever-growing interest of law enforcement authorities and intelligence agencies in the flood of information (cf. [55][56]) from sources as varied as the Web and mobile networks, financial/tax records, travelers' biometric data, satellite imagery, surveillance video, and so forth. Indeed, a clever aggregation and analysis of such data has the potential to gain useful insights into the identities and/or behavioural patterns of (would be) criminals. There is therefore a strong view among law enforcement authorities that the retention and mining of large amount of telecommunication (meta-) data is a key ingredient in resolving crimes committed by means of telecommunication networks or in predicting the possibility of (cyber) attacks on critical infrastructures that are interfacing with Internet nowadays.[57][58] Parallel to the growing adoption of big data storage and analytic techniques to combat threats to organizations' information technology infrastructures and assets, recent years have witnessed the emergence of a different set of big data-based technologies that aimed at law enforcement agencies looking to be able to accurately predict an individual's willingness or readiness to commit a crime, as well as where the next criminal activity might occur. [59] These technologies are enablers of "predictive policing" [60][61][62], a data-driven crime prediction and policing strategy that is increasingly gaining widespread acceptance among police departments across the US [63] and Europe [64]. The predictions about possible criminal activities may be based on the integration of historical crime statistics and external data such as behavioural and content data generated by persons of interest in social media.[65] The underlying methods and tools, however, are similar to statistical tools and advanced computing technologies first deployed by online retailers to predict their customers' behaviour.

---

[53] Gartner Report: Big Data will Revolutionize Cyber Security in the Next Two Years:
http://cloudtimes.org/2014/02/12/gartner-report-big-data-will-revolutionize-the-cybersecurity-in-next-two-year/
[54] Economic Efficiency of Security Breach Notification Analysis and Recommendations.
http://www.enisa.europa.eu/activities/risk-management/files/economic-efficiency-of-security-breach-notification/at_download/file.
[55] The United Kindom Office for Security and Counter Terrorism secret policy for mass surveillance:
https://www.privacyinternational.org/sites/privacyinternational.org/files/downloads/press-releases/witness_st_of_charles_blandford_farr.pdf
[56] http://www.theguardian.com/world/2014/jul/10/surveillance-legislation-commons-support-critics-stitchup
[57] Note that despite his decision on 8 April 2014 declaring the EU's Data Retention Directive invalid, the Court of Justice of the European Union stated that "[... the retention of data for the purpose of their possible transmission to the competent national authorities genuinely satisfies an objective of general interest, namely the fight against serious crime and, ultimately, public security.]" see: http://curia.europa.eu/jcms/upload/docs/application/pdf/2014-04/cp140054en.pdf
[58] Spy Agency Seeks Digital Mosaic to Divine Future: http://www.miller-mccune.com/media/spy-agencyseeks-digital-mosaic-to-divine-future-35878/
[59] Tong Wang, Cynthia Rudin, Daniel Wagner, and Rich Sevieri. Detecting Patterns of Crime with Series Finder. Proceedings of AAAI Late Breaking Track, 2013.
http://www.wired.com/2013/08/predictive-policing-using-machine-learning-to-detect-patterns-of-crime/
[60] Greengard, Samuel. "Policing the future." Communications of the ACM 55.3 (2012): 19-21.
[61] Bachner, J. (2013).Predictive Policing: Preventing Crime with Data and Analytics. IBM Center for the Business of Government.http://www.businessofgovernment.org/sites/default/files/Predictive20Policing.pdf
[62] Rubin, Joel. "Stopping crime before it starts." Los Angeles Times (2010).
[63] Predictive Policing: Preventing Crime with Data and Analytics, http://www.businessofgovernment.org/sites/default/files/Predictive Policing.pdf
[64] http://www.predpol.com/french-media-features-predictive-policing/
[65] Gerber, Matthew S. "Predicting crime using Twitter and kernel density estimation." Decision Support Systems 61 (2014): 115-125.

## 2. Challenges

While the rise of big data yields huge opportunities for individuals, organizations and the society at large, it also raises important privacy and ethical issues. These issues are factors that may lead to situations in which the underlying analytic models and infrastructures are likely to impact privacy negatively from both a legal and an ethical perspective, and hence represent possible obstacles for the big data's potential to be fully realized.

### 2.1. Challenges to Security and Privacy in Big Data

The massive retention of socioeconomic, demographic, behavioural, financial, and other transactional data for analytic purposes may lead to the erosion of civil liberties due to a loss of privacy and individual autonomy. From a privacy and security perspective, the challenge is to ensure that data subjects (i.e., individuals) have sustainable control over their data, to prevent misuse and abuse by data controllers (i.e., big data holders and other third parties), while preserving data utility, i.e., the value of big data for knowledge/ patterns discovery, innovation and economic growth. The following sections describe some relevant challenges to security and privacy in the context of big data.

#### 2.1.1. Increased Potential for Large-scale Theft or Breach of Sensitive Data

As more data is available, stored in (non-) relational databases accessible on-line, and increasingly shared with third parties, the risk of data breaches also increases. Big data thus raises a number of privacy and security questions related to the access, the storage and the usage of personal/ user related data. A recent series of high-profile data security incidents and scandals, e.g. Edward Snowden's NSA leaks [66][67] and the breach at the US retail chain Target Corp [68] have demonstrated that data breaches by those who have obtained access to sensitive datasets, legitimately or otherwise, are devastating for both the individuals and the data holders.

      Unauthorized accesses can possibly involve two types of adversaries: the first type of adversary is interested in gaining access to raw data in order to either compromise the interpretation/analysis process, e.g. by injecting false data into the raw data, or to steal a large volume of sensitive (financial/identity) data. The second type of adversary includes entities primarily interested in accessing different datasets that have already been analysed, as well as the actionable intelligence legitimate analysts have extracted from the data. To breach data privacy, both types of adversaries can exploit software and/or hardware design flaws in the infrastructures behind big data platforms. Thus, the related challenges include preventing security risks and attacks aimed at the underlying big data infrastructure including data centers and cloud platforms where sensitive raw data and inferred knowledge are stored. For individual victims, the consequence of such data breaches is the exposure of sensitive identity attributes and other confidential information (e.g. credit card number) to the greater public. For organizations, data breaches may result in brand damages (i.e., loss of customers and partners trust/loyalty), loss of intellectual property, loss of market share, and legal penalties and fines in case of incompliance with privacy regulations (cf. [69]). For

---

[66] Greenwald, Glenn. No Place to Hide: Edward Snowden, the NSA, and the US Surveillance State. Metropolitan Books, 2014.
[67] The guardian, The NSA Files: Decoded: What the revelations mean for you.
http://www.theguardian.com/world/interactive/2013/nov/01/snowden-nsa-files-surveillance-revelationsdecoded
[68] http://www.bloomberg.com/infographics/2014-05-14/target-data-breach.html
[69] Aimee Picchi. Targets brand takes a massive hit amid data breach. (CBS).

big data holders, the challenging issue is therefore: How to restrict access to, and usage of, sensitive/private data to the authorized parties? Typical solutions to this problem entail encryption, access control, and data anonymization. However, critical data in infrastructures behind big data, e.g., cloud computing, is routinely stored unencrypted. Even when deploying such protection mechanisms, data controllers typically have to rely either on coarse-grained access control models, on encryption schemes that prevent further processing of the data, or on traditional data anonymization which does not work.[70]

### 2.1.2. Loss of Individual Control over Personal Data

As data collection, processing and sharing in the context of big data is becoming more ubiquitous than in traditional online environments, control over personal information flows becomes harder to maintain. The root causes behind this problem include the following:

i) In a typical big data setting, the IT-infrastructure used to collect, store, and process individuals' data as well as the process of drawing inferences from such interconnected data are not under individuals control;

ii) The need for large-scale data storage in the context of big data and fundamental privacy regulatory requirements, including the principles of purpose binding[71] and data minimization[72] are requirements that seem to be conflicting and impossible to fulfill at the same time;

iii) Similarly to other online settings including e-commerce and social media, issues around the ownership of data that is a by-product and knowledge that is inferred through big data analytics are still being discussed controversially. An example showing not only how controversial the issue around data ownership is but also that it remains yet unsolved, is the recent news[73] that telecommunications giants like Verizon, Orange and O2 Telefonica were in the process of beginning selling their consumers' information to buyers looking to use those data e.g., for marketing purposes or managing financial risk. Even when they comply with the law, big data analysts may, similarly to existing data aggregators, do so while relying on poor data anonymization techniques and thus avoiding requesting users informed consent.

These issues do not only raise doubts about whether data subjects right to maintain control over data that they explicitly and implicitly disclose can be fully implemented. They also raise the question whether individual control of personal data throughout the entire data life cycles is a feasible/achievable goal in the context of big data. Another challenging issue closely related to individual control of personal data is the right to access granted to individuals under the European Data Protection Directive. In the face of the complex life-cycle of personal data in big data, the data subjects right to individual access raise an important question - How can it be ensured at the infrastructure level that individuals have real-time access to all of their data (personal data and the inferences drawn from it), request the deletion or update of information they might find uncomfortable, and audit/verify enforcement?

---

http://www.cbsnews.com/news/targets-brand-takes-a-massive-hit-amid-data-breach/
[70] Ohm, Paul. "Broken promises of privacy: Responding to the surprising failure of anonymization." UCLA L. Rev. 57 (2009): 1701.
[71] E. U. Directive,"95/46/EC of the European Parliament and of the Council of 24 October 1995 on the protection of individuals with regard to the processing of personal data and on the free movement of such data." Official Journal of the EC 23.6 (1995).
[72] European Parliament, Council of the European Union: Directive 2002/58/EC on privacy and electronic communications. Official Journal of the European Communities, 31.7.2002, L 201, 3747 (2002)
[73] http://uk.reuters.com/article/2014/02/23/us-mobile-world-bigdata-idUKBREA1M09F20140223

### 2.1.3. Long Term Availability of Sensitive Datasets

The plunging cost of storage coupled with the growing requirements to retain data, both legally and for business purposes and possibly forever, allow governments, private companies and researchers to build big data sets. Besides identity attributes that rarely vary, these big data sets commonly encompass records of an individual's behavioural patterns, and details about an individual's life style, world view, and emotional state of mind. However, the life style, habits, and world view of an individual usually change several times in her or his life. Meanwhile, the digitalization of our society has changed the discussion about the availability of (personal) digital information forever: information is easily reproducible, mostly stored in cross-linked databases that are searchable instantly, and deleting that information is often, not a practical expectation in the online world as we know it. Big data sets (including knowledge inferred through analysis) may therefore include permanent records of all the disclosures and mistakes the data subject has ever made online and would not like anybody to remember them years afterwards. This calls on data analysts for more consideration of long-term security risks and life-long privacy management of data records which data analysts may intend to store and use for a very long period (10-30 years or even longer).

### 2.1.4. Data Quality/Integrity and Provenance Issues

As big-data-enabled applications are information rich and context sensitive per se, and often require knowledge of the history and genealogy of the data objects being considered, addressing the issues of data quality, data provenance and data integrity in the presence of potentially untrusted data sources is becoming a major challenge faced by users of big data analytics (i.e. companies, research and governments). Without means to maintain the integrity and the quality of the data on the one hand, and to capture and understand details about both the data's pedigree and the context in which it was collected, big data analysts will have a hard time achieving business optimization, managing their mountain of data or running any valuable data-driven process/operation. Indeed, it is the details about the quality, the provenance and the integrity of the data that in large part determines how users of big data analytics should analyze the data and how they should interpret the results of the analysis, especially in context-sensitive settings such as data dependency analysis, strategic and tactical decision optimization within organizations, and malicious/criminal behaviour detection by law enforcement authorities. According to a 2011 survey by Gartner [74], 40% of business initiatives fail due to poor data quality. This indicates, especially in the context of big data, the need for a carefully designed data collection process as well as additional measures to reduce/avoid the risk of wrong decisions and misinterpretation that are due to inaccurate data. Acting based on insights gained from either poor/inaccurate data or imprecise big data models may have serious, negative implications for the data subject including being sorted into categories, which often reinforce existing social stereotypes, subsequently being subjected to unfair / unlawful discrimination, denied access to services (because the city may have decided to restrict investment in poor neighbourhoods) or otherwise subjected to unpleasant experiments.[75]

---

[74] Ted Friedman, Michael Smith. Measuring the Business Value of Data Quality. Gartner. 10 Oct. 2011.
[75] The data subject may not get the job she wants only because she live in a particular city area, her insurance premium may go up, law enforcement may decide to "stop and frisk" her on the street or she may not gain access to the education of her choosing.

## 2.1.5. Unwanted Data Correlation and Inferences

Aggregating and cross-linking large volume of data produced by and about people, and coming from heterogeneous sources in order to discover hidden patterns raise additional risks as big data holds the potential to accentuate the privacy implications of database correlations (cf. [76]) dramatically. Indeed, the correlation of information from diverse data sources, a typical operation in most big data settings, may increase the risk of re-identification / de-anonymization.[77][78][79] Considered separately, some pieces of data from various sources might be innocuous-seeming data or even "de-identified datasets", i.e., "free" from identifiable information. Data analyst may therefore tend to consider them as not covered by existing data protection laws. However, analysts[80] can intentionally or accidentally, identify new inferences or discover new sets of sensitive information the data subject has not agreed to share. Analysts can do this by correlating supposedly "de-identified" data sets with publicly or other privately available data sets. [81][82] [83] Using the inferred information/knowledge, dishonest analysts can make legal but unfair decisions causing substantial damages or distresses to people whose data is stored and correlated. Note however that privacy issues as consequences from correlating pieces of public data are not new [84][85]. Several years ago, Latanya Sweeney [86][87] showed that it is possible to retrieve personal health data by cross-linking an allegedly anonymous database of Massachusetts state employee health insurance claims with a publicly available voter registration database. In 2009, Acquisti and Gross [88] demonstrated that one can predict an individual's social security number with remarkable accuracy using only public data. For their test, the scientists used the publicly available U.S. Social Security Administration's Death Master File and other personal data from multiple sources, such as data brokers or profiles on social networking sites. In 2012 Charles Duhigg, a staff writer for the New York Times and author of "The Power of Habit: Why We Do WhatWe Do in Life and Business" [89] described in a piece [90] how statisticians for Target[91] are able to predict a female customer's pregnancy (with an 87 percent confidence rate) by tracking all customers' purchases histories and identifying patterns in their behaviour. Using the inferred insights, target was subsequently able

---

[76] Csilla Farkas and Sushil Jajodia. 2002. The inference problem: a survey. SIGKDD Explor. Newsl.4, 2 (December 2002), 6-11. DOI=10.1145/772862.772864 http://doi.acm.org/10.1145/772862.772864

[77] Narayanan, Arvind, and Vitaly Shmatikov. "De-anonymizing social networks." Security and Privacy, 2009 30th IEEE Symposium on. IEEE, 2009. http://dx.doi.org/10.1109/SP.2009.22

[78] Narayanan, Arvind, et al. "On the feasibility of internet-scale author identification." Security and Privacy (SP), 2012 IEEE Symposium on. IEEE, 2012. https://www.ieee-security.org/TC/SP2012/papers/4681a300.pdf

[79] Goga, Oana, et al. "Exploiting innocuous activity for correlating users across sites." Proceedings of the 22nd international conference onWorldWideWeb. InternationalWorldWideWeb Conferences Steering Committee, 2013. http://www2013.wwwconference.org/proceedings/p447.pdf

[80] For instance, those working for Internet giants or state surveillance entities with increase access to ever growing storage, and advanced processing capabilities.

[81] Matching Known Patients to Health Records in Washington State Data. Harvard University. Data Privacy Lab. 1089-1. June 2013.

[82] Narayanan, Arvind, and Vitaly Shmatikov. "Myths and fallacies of personally identifiable information." Communications of the ACM 53.6 (2010): 24-26. Sweeney L.

[83] Ohm, Paul. "Broken promises of privacy: Responding to the surprising failure of anonymization." UCLA L. Rev. 57 (2009): 1701.

[84] Sweeney, Latanya. "Simple demographics often identify people uniquely."Health (San Francisco) (2000): 1-34. http://dataprivacylab.org/projects/identifiability/paper1.pdf

[85] Sweeney, Latanya. "k-anonymity: A model for protecting privacy." International Journal of Uncertainty, Fuzziness and Knowledge-Based Systems 10.05 (2002): 557-570. http://www.worldscientific.com/doi/abs/10.1142/S0218488502001648

[86] Sweeney, Latanya. "Simple demographics often identify people uniquely."Health (San Francisco) (2000): 1-34. http://dataprivacylab.org/projects/identifiability/paper1.pdf

[87] Sweeney, Latanya. "k-anonymity: A model for protecting privacy." International Journal of Uncertainty, Fuzziness and Knowledge-Based Systems 10.05 (2002): 557-570. http://www.worldscientific.com/doi/abs/10.1142/S0218488502001648

[88] 11. "Predicting Social Security Numbers from Public Data" Alessandro Acquisti and Ralph Gross. Proceedings of the National Academy of Science, 106(27), 10975-10980, 2009.

[89] Kirshner, Howard S. "The Power of Habit: WhyWe Do WhatWe Do in Life and Business." (2013): 50-51.

[90] http://www.nytimes.com/2012/02/19/magazine/shopping-habits.html?pagewanted=1

[91] The Target Corporation (http://www.target.com/) is one the largest retailing company in the USA

to serve those customers identified as expectant parents personalized advertisements about new born related goods. More recently, a study by Kosinski et al. [92] has shown that using Facebook Likes, it is possible to infer automatically and accurately sensitive identity attributes of social media users including age, gender, sexual orientation, ethnicity, religious and political views, and other personality traits. These and similar researches (e.g., [93][94][95][96]) demonstrate the power of data correlation and inference attacks, on the one hand, and suggest that big data could make things worse, on the other hand. The rises of large data pools produced by and about people, that interact with each other, combined with the emergence of powerful data mining techniques, clearly increase the potential for privacy violations, e.g. re-identification, identity theft, and discrimination against individuals or minorities.

### 2.1.6. Lack of Transparency and (the Limits of) Consent Management

Electronic consent and the closely related notion of "notice" are two key legal requirements imposed upon entities (private/public organizations and individuals) that collect, process and share personally identifiable information (PII). Both concepts are an integral part of most privacy and data protection legislations/regulations around the world. However, depending on the jurisdiction a data controller is subject to, the notion of consent may have different definitions and be subject to different interpretations by legislators, privacy theorists and advocates, citizens or law enforcement authorities. For example, while the consent of data subject is understood in most regulations in North America as a contractual act, i.e. reflecting the idea of an contractual relationship between a data subject and a data controller, privacy directives in Europe see consent as an unilateral act, i.e. a form of single manifestation of the data subject's will [97]. Under the EU Data Protection Directive 95/46/EC [98] the data subject's consent is specified as " […] any freely given, specific and informed indication of his wishes by which the data subject signifies his agreement to personal data relating to him being processed". This implies that the data subject has been offered a meaningful choice of options as well as the opportunity to evaluate both the benefits and potential risks of its choice. Accordingly, consent given in a situation with few realistic alternatives hardly fits within the notion of freely given consent. Further, a data subject's consent must have not been given under pressure or on the basis of misleading information. Hence, the data subject must always be able to (temporarily) withhold her/his consent.

Implementing these regulatory requirements in big data settings is a challenging task since it would require a shift from empowering individuals at the point of data collection, which traditionally includes opting out/ into privacy notices, to empowering them to make informed consent decisions at processing time. Yet current approaches for requesting and managing consent remain questionable practices [99] - individuals tend to paid limited

---

attention to notices (Privacy policies on the bottom of website, EULA, permission screens) (cf. [100][101][102]), consent to whatever terms when using/installing digital services, and have poor understanding of what consequences their actions (e.g., agreeing to certain present terms) implied -, which the dynamic nature of big data processing might further aggravate. Indeed, as the types of inferred information, the possible purposes for which and contexts in which this information can be used, and the kinds of predictions that can be made from it are not always perceptible at data collection time, people cannot be expected to be able to consent to the use thereof beforehand. Moreover, the type and value of information inferable using big data techniques, the possible purposes and contexts for secondary use of individuals personal data are far too numerous and constantly evolving for users to be able to adequately assess potential consequences when making a cost-benefit analysis over whether to consent to the collection, use, or disclosure of their personal data. Another challenging issue arising from applying powerful analytics on the aggregation of (non-) sensitive individuals' data is the power/information asymmetry between data subject and data controllers (see section 2.2.). Such an asymmetry does not only shape the way privacy choices are framed, it also further limit individuals' ability to understand the complex ramifications of their consent decisions. In addition, acknowledging the fact that individuals consent decisions do not only impact themselves but rather have consequences for others both on- and offline, it is difficult to image how individuals in big data settings would make informed decisions about whether to consent to the disclosure of data without having enough expertise and accurate knowledge of the broader set of community values and expectations about privacy. Existing approaches to notice and consent neglects to address this issue.

As a matter of fact, and given that the infrastructures and big data algorithms used to collect and process individuals' data may operate opaquely, out of sight of the data subjects, i. e, data subjects may not always be aware of the scale of such infrastructures, or how the algorithms work precisely, a meaningful implementation of the concepts of notice and consent management in big data requires reliable enforcement of consent decisions and empowering individuals with transparency, both ex-ante and ex-post transparency. Indeed, the data subject need on the one hand to be provided, prior to the consent decision, with relevant insight about data controllers' privacy practices and potential implications/ risks of her choice (ex-ante transparency). On the other hand, there can be concerns about the lack of transparency towards the data subject about when, and how the subject's personal data is / has been manipulated (ex-ante transparency), especially when it comes to scenarios in which individuals' data may has been share and process across various services domains and for purposes other than those identified at the time the information was collected. Given the sheer volume and heterogeneity of data as well as the complexity of the data analysis, it is unclear how to ensure transparency towards the data subject about how this information is being collected and used/interpreted, where it is currently located/ processed, and which new knowledge has been inferred from it, for what purposes. Indeed, it can be difficult after multiple aggregation/combination of different datasets to determine the pedigree of any a single piece of data as required by law (cf. [103]) and to link that data to a particular person.

---

So the transparency challenge in big data settings boils down to one question: How does one provide data subjects the ability to i) be aware of and track complex flows and use of personal data, ii) better understand risks and harms this data may enable, and iii) make disclosure choices and decisions that reflect their privacy preferences? Furthermore, when pulling data from different sources and using it to take advantage of the opportunity offered by Big data (see section 1), one concern is that personal information torrent from external sources may have been collected and shared without users' informed consents. This raises questions regarding the possible infringements of users' right to self-determination and big data holders' ability to maintain regulatory compliance. Data controllers have argued that the concept of informed consent sit uneasily with modern data-driven business models: requesting and managing consent is often associated with additional logistical and financial costs.

As such, these limitations call for a new approach to transparency and consent management aiming to better support consent specification (i.e. signaling of consent decision), refinement (i.e., to support alternatives to current all-or-nothing approach that poorly fits users' needs), enforcement, and revocation in the emerging big data-driven businesses models.

### 2.1.7. Algorithmic Accountability

In addition to transparency, algorithmic accountability is another challenging issue that arises from the fact that big data algorithms while being enablers of societally beneficial goals such as those discussed in section 1 might also have negative implications for individuals, including ubiquitous profiling and surveillance and various forms of discrimination (see section 2.2.). Addressing these issues would require moving beyond the current approach to accountability i.e. empowering data subjects (and possibly Data protection authorities - DPA) with the ability to check whether a computation on their data was performed accurately and according to their privacy preferences and existing norms. In addition, data subjects and DPA should be able to check whether algorithms have been used improperly, i.e., whether the data controllers' decision, in a particular context, based on inferred associations are deemed acceptable/fair. However, while the ultimate decision to use a certain inference in a certain way may be made by humans (data analysts), the algorithms used to mine sets of data and discover the inference are often challenging to understand for many - they are often too complex and opaque.

### 2.2. Ethical and Social Challenges

In addition to the legal and technical challenges discussed in the section above, big data technology raises critical ethical and moral issues that society has yet to fully address. Below, we list some of these issues.

### 2.2.1. Information Asymmetry and the Issue of Power

In economic terms, the term "information asymmetry" describes a market situation in which one of the parties involved in a transaction possesses superior information [104][105]. A typical example is the situation in the retail industry where the retailer usually has more and/or better information than its buyer since it can rely on accurate information about the availability of goods, knowledge of the exact prices charged by suppliers, and its experience over the years. In contrast, the buyer's information is limited and often based on only few purchase experiences (including those shared by peers). A similar situation exists between a provider (lender) and a consumer (borrower) of financial products. As related to big data, information asymmetry could allow a powerful few[106] to access and use knowledge that the majority, including individuals, do not really have access to - leading to a (or exacerbating the existing) asymmetry of power between the state (resp. big business) and the people (resp. customers). The ability to accumulate and manipulate data about customers and citizens on an unprecedented scale may give big companies with selfish agendas and intrusive/authoritarian governments powerful means to manipulate segments of the population through targeted marketing efforts, perform social control, and hence possibly negatively impact the course of our democracies, or do all sorts of harm. (see section 2.2.2 and section 2.2.3) Moreover, the increasing monolithic power which users of big data analytics - like Internet giants Google, Facebook, or Amazon - derive from the data that they hold also raise challenges for information transparency and informational self-determination. Indeed, big data holders usually do not reveal in clear terms which of the individuals' data they exactly collect, and for what purpose they used it. (see section 2.2.1) On the other hand, even when assuming that the big data holders would provide this information, individuals may still lack the capability to fully understand it and/or to make informed decisions on this basis. (cf. [107]) Moreover, reflection about information asymmetry and the related issue of power inevitably lead to concerns about unlawful, pervasive off-/online surveillance, perpetuation of existing forms of social stereotypes, and unfair discrimination.

---
[104] Akerlof, G.A.: The market for lemons: Quality uncertainty and the market mechanism. Quarterly Journal of Economics 84(3) (1970) 488500
[105] Ronald J. Mann, Verification Institutions in Financing Transactions, 87 GEO. L.J. 2225, 2248-49 (1998).

### 2.2.2. Surveillance

In a trend very similar to George Orwell's science fiction "Nineteen Eighty-Four" and Philip K. Dick's "Minority Report", big data technologies are used increasingly for ubiquitous, systematic, large-scale off-/online surveillance, not just by intelligence and law enforcement agencies [108], but also by commercial entities [109][110] and individuals themselves [111][112]. The infrastructures and big data algorithms used by online retailers and other service providers to collect and analyze large volumes of internet user-related data might not only help them to understand customer needs better, and subsequently provide more personalized services and better-targeted advertising. It also helps them to build a treasure trove of digital, personal dossiers that include details like browsing and spending habits, online social interactions – people we are connected to online, topics of our conversations, places we have visited, activities inside homes, our daily routine, search terms, inclination toward a brand, and so forth. Leveraging such personal dossiers, privacy-invasive entities (predomi-

---

[106] In [*Boyd, Danah, and Kate Crawford. "Critical questions for big data: Provocations for a cultural, technological, and scholarly phenomenon." Information, Communication Society 15.5 (2012): 662-679.*] Boyd and Crawford refer to a new kind of digital divide, i.e., the one between the big data rich (company researchers that can purchase and analyse large datasets) and the big data poor (those excluded from access to the data, expertise, and processing power), highlighting the fact that the "the Big Data rich", a relatively small and most privileged group, is getting control over how big data will be used for research purposes.
[107] Solove, Daniel J., Privacy and Power: Computer Databases and Metaphors for Information Privacy. Stanford Law Review, Vol. 53, p. 1393, July 2001. Available at SSRN: http://ssrn.com/abstract=248300 or http://dx.doi.org/10.2139/ssrn.248300
[108] Greenwald, Glenn. No Place to Hide: Edward Snowden, the NSA, and the US Surveillance State. Metropolitan Books, 2014.
[109] What They Know series by the Wall Street Journal, http://online.wsj.com/public/page/what-they-knowdigital-privacy.html.
[110] Markus Schneider, Matthias Enzmann, Martin Stopczynski. Web-Tracking-Report 2014. Darmstadt Fraunhofer SIT / Michael Waidner (Herausgeber) 2014
[111] Mann, Steve, Jason Nolan, and Barry Wellman. "Sousveillance: Inventing and Using Wearable Computing Devices for Data Collection in Surveillance Environments." Surveillance Society 1.3 (2002): 331-355.
[112] Self-surveillance privacy

nantly corporations and government spy agencies) may infer sensitive private attributes such as religious/political views and monitor peoples' online behaviours in real-time. Additionally to this kind of profiling, predictive models used to gain insights about market, economic or societal trends may be exploited to create global-scale systems of ubiquitous surveillance, which, supposedly for the purpose of national security and to maintain public order, tend to target everybody – not only individuals, but groups and organizations as well that the state deems subversive and dangerous - both on- and offline. Evidences from the Snowden revelations [113] and details about existing predictive policing programs [114] suggest that big-data-based national security and law enforcement instruments have been turned into surveillance apparatuses, leading to a situation in which fundamental principles like the presumption of innocence are questioned: everybody is presumed guilty before proven innocent. Moreover, whenever data generated by or related to people is collected, and stored systematically, the expected result is large information repositories about various aspects of people's life that government agencies will be eager to gain access to, e.g., by means of subpoenas. [115] [116]

Despite existing data protection frameworks[117], which guarantee the right not to be subjected to such surveillance, big data analytics can expand governments' and businesses' ability to surveil private citizens, ultimately undermining individuals' right to privacy, anonymity, and free expression. Without consensus in our society on the scope of data retention and surveillance laws and their interplay with data protection laws, big data programs risk being negatively impacted, derailed or even stopped by public backlash and legal challenges. [118][119]

### 2.2.3. Filter Bubble, Social Sorting, and Social Control: By-products of Unfair Discrimination

According to conventional wisdom, minorities and the powerless need protection and solidarity. However, big data analytics accessible only to the powerful few raise questions not only about privacy and security but also about fairness and social sorting. Indeed, big data increases the chance, the accuracy, and the ease with which unfair discrimination directed at certain individuals might occur. While the growing ability to collect and analyze vast amounts of data, may reveal high value insights used to improve decision making and operational efficiency, and deliver services tailored to people's needs, it also allows the discovery of knowledge that might be used for discrimination. Indeed, when deployed in scenarios such as price discrimination [120] and for other purposes such as consumer scoring and law enforcement (e.g., predictive policing), advanced mathematical models of personal and group behaviours can be exploited in ways that treat people unfairly and are even prohibited (cf. [121][122][123]). The issue here being that these models may not only help analysts to understand the similarities and differences between individuals better, but also provide them with means to segregate individuals into various categories and groups. Those targeted (e.g. single person or ethnic minority groups) may suffer negative consequences[124] while lacking both the knowledge of the fact that they have been discriminated against and the ability to properly defend themselves. In the following, a few examples of anecdotes: Insurance companies and healthcare providers able to collect and analyze information from public health repositories, poorly de-identified patient databases, disease-related web search logs, or patients' online communities can discover insights, which they might use not only to optimize their services' quality and efficiency, but also to profile individuals, reverse identify, and then discriminate against specific individuals on the basis of sexual orientation or gender identity.

---

[113] Greenwald, Glenn. No Place to Hide: Edward Snowden, the NSA, and the US Surveillance State. Metropolitan Books, 2014.
[114] Jennifer Bachner. Predictive Policing: Preventing Crime with Data and Analytics, The IBM Center for The Business of Government, 2013.
[115] https://www.google.com/transparencyreport/userdatarequests/
[116] https://govtrequests.facebook.com/
[117] Mostly across western democracies: see Art. 15 jo 12 D 95/46 EC and art. 20 of the proposed General Data Protection Reform (GDPR) in Europe; The Communications Assistance for Law Enforcement Act (CALEA) in the U.S.
[118] Konstadinides, Theodore. "Mass Surveillance and Data Protection in EU Law: The Data Retention Directive Saga." (2012).
[119] Court of Justice of the European Union. PRESS RELEASE No 54/14. Luxembourg, 8 April 2014.
Judgment in Joined Cases C-293/12 and C-594/12 Digital Rights Ireland and Seitlinger and Others.
http://curia.europa.eu/jcms/upload/docs/application/pdf/2014-04/cp140054en.pdf
[120] Personalized coupons as a vehicle for perfect price discrimination. http://33bits.org/tag/pricediscrimination/
[121] Dixon, Pam and Gellman, Robert. The Scoring of America: How Secret Consumer Scores Threaten Your Privacy and Your Future, World Privacy Forum, 2014.
[122] Michel Waelbroeck, Price Discrimination and Rebate Policies under EU Competition Law,
 (1995) Fordham Corporate Law Institute, 148.

In a recent New York Times article [[125]], M. Hudgins reported on how specialized software is allowing real estate firms to calculate and suggest automatically the highest possible rent price to charge for a particular apartment by analyzing mountains of information including real estate supply and demand statistics, market prices, marketing surveys, and government censuses. Some have expressed concerns that this practice may lead to an increase in housing discrimination, i.e. it would make it difficult for certain classes of citizens (e.g. low-income families, seniors, and people with disabilities) to find affordable rental housing. While big data proponents may argue that this will help landlords reduce the risk of letting to unreliable tenants, one can also imagine sensitive inferences and facts about possible tenants being used in ways that contribute to unfair discrimination and perpetuate economic and social inequality.[[126]] In addition, combining data about properties and government released crime statistics, and plotting them on a map may lead to discriminate against poor neighborhoods and negatively affect the property values. Airline companies and hotel chains are using similar techniques to collect and analyze vast amounts of information about available seats or rooms, marginal costs, and expected demand. Based on the findings the advanced analysis revealed, the companies can vary their prices profitably. [[127]] In these scenarios, big data models are used to estimate an individual's willingness to pay a certain monetary price for a service or to compute and assign a credit worthiness score to customers in both cases based on large sets of personal information, including race, religion, age, gender, household income, ZIP code, medical conditions and purchase habits. In the process, the population/ customer base is segmented into specific social groups that are "treated" differently. For instance, a person might be proposed a lower credit limit, mostly because the inferred patterns suggest that people from her neighborhood have a bad repayment history.

When applied in the context of law enforcement, big data analytics can encourage similar segmentation of our society, raising serious civil rights concerns, as inferences and correlations (accurate or not) can help perpetuate negative social stereotypes (e.g., racial profiling as a dark side of preventive policing). In the realm of politics, evidence from the U.S. elec-tions in 2008 and 2012 suggest that people are becoming increasingly comfortable with disclosing personal information (including details about their social ties both on and offline) in exchange of a highly personalized political experience tailored to their own views and needs. Such a trend, however, raises any number of civil rights concerns, including the risk to aggravate the increasing ideological uniformity and partisan antipathy (cf. [128]), thus treat the very fabric of our democratic and free society. Moreover, given the lack of transparency about what is being done with (inferred from) individuals data, coupled with the fact that big data models (or interpretation of the results) may be flawed [129], the sorting and targeting of individuals based on predictions or inferences may result in forms of guilt-by-association. For instance, an insurance company with access to genetic information about one of its client may use it to predict health issues the client might face in the future, and then deny claims, or even insurance coverage not only for that particular client, but also the client's relatives. Another combined effect of such an (controversial / unlawful) use of big data goes further behind aggravating the economic and cultural divides that already shape so many part of our society. Indeed, big data-driven forms of social sorting and power asymmetry may facilitate various forms of social control. That is, individuals in a big data enabled surveillance society may tend, similarly to the prisoners in Jeremy Bentham's "Panopticon" [130], to avoid behaving or expressing points of view that could put them under further scrutiny without due process or the ability to defend themselves properly (since people usually have few effective ways to challenge unfair use of sensitive statistical inferences [131]).

---

[123] Helen Jung. No-fly list appeals process unconstitutional, federal judge in Portland rules.
http://www.oregonlive.com/portland/index.ssf/2014/06/no_fly_list_appeals_process_un.html
[124] Note that automated decision making may facilitate the sorting of people into categories. In certain categories, individuals may be subjected to even further (unfair) scrutiny resulting in loss of autonomy and reputation.
[125] M. Hudgins. When Apartment Rents Climb, Landlords Can Say 'The Computer Did It. November 29, 2011
[126] Steering 2.0? Data may undermine fair housing laws. http://www.inman.com/2014/04/29/steering-2-0- data-may-undermine-fair-housing-laws/.U3l7W-ZdVv8
[127] Price Discrimination By Day-Of-Week Of Purchase: Evidence From The U.S. Airline Industry [PDF], published in the Journal of Economic Behavior Organization

## 3. Conclusion

The collection and analysis of large volume of data is not a new concept. In fact, large organizations in industries like banking and finance have used complex algorithms and tools for years to handle and extract meaning from large quantities of structured data. Nowadays, however, data analysts from various sectors of the society are increasingly interested in collecting, and analyzing new types and sources of under-leveraged data, e.g., sensors, social media, emails, and mobile network data. As the capability of analysts to leverage recent advancements in analytics and high-performance computing is improving, so is their ability to extract valuable insights from huge volumes of complex and heterogeneous data that would otherwise remain hidden.
Despite being viewed as an enabler of breakthroughs in key sectors of society, such as healthcare, science, business, law enforcement and national security, big data analytics entail important privacy, security and ethical challenges that technologists, regulators, business and the society at large have yet to address.
This paper presented an overview of some key privacy and ethical issues that accompany the rise of big data. Our work is aimed at developing a better understanding of these issues. Future work, which is already underway, is the investigation of approaches to built trust and privacy adequately into big data technology, focusing on striking the delicate balance be-tween the people's right to privacy and the need to extract additional knowledge, patterns, and value from (personal) data.

---

[128] http://www.people-press.org/2014/06/12/political-polarization-in-the-american-public/
[129] Batler, D. When Google got flu wrong. US outbreaks foxes a leading web-based method for tracking seasonal flu. http://www.nature.com/news/when-google-g?ot-flu-wrong-1.12413.
[130] Bentham, Jeremy. "Panopticon, or The Inspection House,(1787)." The Panopticon Writings: 29-95.
[131] For instance when put on a No Fly List as consequences of points of view expressed in the past or of being member of an ethnic minority group which the state deemed deviant/ dangerous. http://www.washingtonpost.com/world/national-security/lawsuit-alleges-fbi-is-using-no-fly-list-to-force-muslims-to-become-informants/2014/04/22/1a62f566-ca27-11e3-a75e-463587891b57_story.html

## Acknowledgements

This work has been supported in part by the German Federal Ministry of Education and Research (BMBF) within Forum Privatheit (http://www.forum-privatheit.de/forum-privatheit-de/) and EC SPRIDE (www.ec-spride.de).